\documentclass[
twocolumn,
preprintnumbers,
nofootinbib,
floats,
prl,
aps
]{revtex4-1}

\usepackage{graphicx}
\usepackage{amsmath, amssymb}
\usepackage{slashed}
\usepackage[T1]{fontenc}

\begin{document}

\preprint{DESY 14-237}

\title{Grand Unification and Subcritical Hybrid Inflation}

\author{Wilfried Buchm\"{u}ller and Koji Ishiwata}

\affiliation{ Deutsches Elektronen-Synchrotron DESY, Notkestrasse 85,
  22607 Hamburg, Germany }
 
%\date{\today}

\begin{abstract}
\noindent
We consider hybrid inflation for small couplings of the inflaton to
matter such that the critical value of the inflaton field exceeds the
Planck mass. It has recently been shown that inflation then continues
at subcritical inflaton field values where quantum fluctuations
generate an effective inflaton mass. The effective inflaton potential
interpolates between a quadratic potential at small field values and a
plateau at large field values. An analysis of the allowed parameter
space leads to predictions for the scalar spectral index $n_s$ and the
tensor-to-scalar ratio $r$ similar to those of natural
inflation. Using the ranges for $n_s$ and $r$ favoured by the Planck
data, we find that the energy scale of the plateau is constrained to
the interval $(1.6 - 2.4)\times 10^{16}\,\mathrm{GeV}$, which includes
the energy scale of gauge coupling unification in the supersymmetric
standard model.  The tensor-to-scalar ratio is predicted to satisfy
the lower bound $r > 0.049$ for $60$ $e$-folds before the end of
inflation.
\end{abstract}

\maketitle

\section{I. introduction}

The observations and analyses of the cosmic microwave background (CMB)
by the WMAP \cite{Hinshaw:2012aka} and Planck \cite{Ade:2013uln}
collaborations strongly support single-field slow-roll inflation as
the paradigm of early universe cosmology. The current CMB data can be
successfully described by many models of inflation. Prominent examples
are the Starobinsky model \cite{Starobinsky:1980te}, chaotic inflation
\cite{Linde:1983gd}, natural inflation \cite{Freese:1990rb} and hybrid
inflation \cite{Linde:1993cn}, which differ significantly in their
predictions for the scalar spectral index $n_s$ and the
tensor-to-scalar ratio $r$ of the primordial density fluctuations. The
recently released BICEP2 data \cite{Ade:2014xna}, which are presently
under intense scrutiny
\cite{Mortonson:2014bja,Flauger:2014qra,Adam:2014bub}, have renewed
the interest in models with a large fraction of tensor modes.

A theoretically attractive framework is supersymmetric D-term
inflation \cite{Binetruy:1996xj,Halyo:1996pp,Kallosh:2003ux}. It is
remarkable that it contains the rather different models listed above
for different choices of the K\"ahler potential: For a canonical
K\"ahler potential one obtains standard hybrid inflation, for a
superconformal or no-scale K\"ahler potential the Starobinsky model
emerges \cite{Buchmuller:2012ex}, and in case of a shift symmetric
K\"ahler potential \cite{Kawasaki:2000yn} D-term inflation includes a
``chaotic regime'' with a large tensor-to-scalar ratio
\cite{Buchmuller:2014rfa}.

In this note we study the chaotic regime of D-term inflation in more
detail. It turns out that the predictions are qualitatively similar to
those of natural inflation, although the theoretical interpretation is
entirely different. Moreover,  there are significant quantitative
differences.  

The parameters of D-term inflation are a Yukawa coupling, the
Fayet-Iliopoulos (FI) term and a gauge coupling. The last two
determine the energy scale $M_{\rm inf}$ of hybrid inflation.  The
measured amplitude of scalar fluctuations determines $M_{\rm inf}$ as
function of the Yukawa coupling.  Imposing the bounds of the Planck
data on $n_s$ and $r$ as constraints \cite{Ade:2013uln},
\begin{equation}
\begin{split}\label{2sigma}
n_s & = 0.9603 \pm 0.0073\ , \\
r &  <  0.11  \,(95\%\,{\rm CL})\ ,
\end{split}
\end{equation}
we find that $M_{\rm inf}$ has to be close to the energy scale $M_{\rm
  GUT}$ of grand unification. Furthermore, we obtain a lower bound on
the tensor-to-scalar ratio, $r > 0.049\,(0.085)$ for $60\,(50)$
$e$-folds before the end of inflation, which is in reach of upcoming
experiments.

\section{II. Subcritical Hybrid Inflation}

The framework of D-term hybrid inflation in supergravity is defined by
a K\"ahler potential, a superpotential and a D-term scalar potential
\cite{Binetruy:1996xj,Halyo:1996pp,Kawasaki:2000yn,Buchmuller:2014rfa},
\begin{align}
K &= \frac{1}{2}(\Phi+\Phi^{\dagger})^2 
+ |S_+|^2 + |S_-|^2\ ,
\\
W &= \lambda\ \Phi S_+ S_-\ , \\
V_D &= \frac{g^2}{2} \left(|S_+|^2 - |S_-|^2 - \xi\right)^2 \ .
\end{align}
The ``waterfall fields'' $S_\pm$ carry the U(1) charges $\pm 1$, and
the inflaton is contained in the gauge singlet $\Phi$. The K\"ahler
potential is invariant under the shift ${\rm Im}(\Phi) \rightarrow
{\rm Im}(\Phi) +\alpha$ where $\alpha$ is a real constant, i.e., it is
independent of the constant part of ${\varphi}\equiv \sqrt{2}\,{\rm
  Im}(\Phi)$, which is identified as the inflaton field. The gauge
coupling $g = \mathcal{O}(1)$, and $\lambda$ is a Yukawa coupling,
which may be much smaller than $g$. The only dimensionful parameter is
the FI term $\xi$ that sets the energy scale of
inflation.\footnote{Note that FI terms in supergravity are a subtle
  issue
  \cite{Binetruy:2004hh,Komargodski:2010rb,Dienes:2009td,Catino:2011mu}. For
  recent discussions and references on field-dependent and
  field-independent FI terms, see
  refs.~\cite{Wieck:2014xxa,Domcke:2014zqa}. In the following we shall
  treat $\xi$ as a constant.}

Standard hybrid inflation takes place at inflaton field values
$\varphi$ larger than the critical value $\varphi_c=(g/\lambda)
\sqrt{2\xi}$. Here the waterfall fields $S_\pm$ have a positive mass
squared and are stabilized at the origin. Classically, the potential
is independent of modulus and phase of the gauge-singlet $\Phi$. The
flatness in $|\Phi|$ is lifted by quantum corrections.

For subcritical field values $|\Phi| < \sqrt{2} \varphi_c$ the
complex scalar $S_-$ remains stabilized at the origin, whereas 
$S_+$ acquires a tachyonic instability. The sum of F- and D-terms
yields for the scalar potential as function of $\varphi$ and
$s\equiv\sqrt{2}|S_+|$,
\begin{align}\label{eq:Vtot}
V(\varphi,s) &= V_F(\varphi,s) + V_D(s) \\
&= \frac{\lambda^2}{4} s^2\varphi^2
+\frac{g^2}{8}\left(s^2-2\xi\right)^2+
\mathcal{O}(s^{2n}\varphi^2) \ , \ n\geq 2\ . \nonumber
\end{align}
Note that, due to the shift symmetry of the K\"ahler potential,
the Planck suppressed terms are also only quadratic in
$\varphi$. The scalar potential contains
higher powers in ${\rm Re}(\Phi)$, which we have neglected since they
are not important for inflation. 

Following ref.\,\cite{Buchmuller:2014rfa}, we solve the classical
equations of motion for homogeneous fields, corresponding to the
scalar potential \eqref{eq:Vtot},
\begin{equation}\label{classical}
\begin{split}
\ddot \varphi + 3H\dot \varphi + \frac{\lambda^2}{2}  s^2 \varphi &= 0 \ ,  \\
\ddot s + 3H\dot s - \left(g^2\xi - \frac{\lambda^2}{2}\varphi^2\right)s +
\frac{g^2}{2}s^3 &= 0 \ .
\end{split}
\end{equation}
The initial conditions for the waterfall field are obtained by
considering the tachyonic growth of its quantum fluctuations
\cite{Felder:2000hj,Asaka:2001ez,Copeland:2002ku,Dufaux:2010cf} close
to the critical point $\varphi_c$,
\begin{align}\label{growth}
 \langle s^2(t) \rangle \simeq 
\int_0^{k_b(t)} dk \,\frac{k^2}{2 \pi^2}\, e^{- 3 H_c t}\, | s_k(t)|^2
\ .
\end{align}
Here $s_k(t)$ are the momentum modes of the field operator in an
exponentially expanding, spacially flat background with Hubble
parameter $H_c = H(\varphi_c)$ and a time-dependent inflaton field
$\varphi(t) = \varphi_c + \dot\varphi_c t$ \cite{Asaka:2001ez}, 
\begin{align}\label{modes}
\ddot{s}_k + \left(k^2e^{-2H_ct}-\frac{9}{4}H_c^2-D^3t\right)s_k=0\ .
\end{align}
The integration in eq.\,\eqref{growth} extends over all soft momentum
modes below $k_b(t)$ where the time-dependent mass operator for
$s_k(t)$ in the brackets of eq.\,\eqref{modes} vanishes. At a
decoherence time $t_{\rm dec} \sim (3\ln(2 R_{\rm dec})/4)^{2/3}/D$,
where $R_{\rm dec} \sim 100$ and $D = (\sqrt{2 \xi} g \lambda |\dot
\varphi_c|)^{1/3}$, the waterfall field becomes classical. Matching
the variance and classical field near the decoherence time, $s(t)
\equiv \langle s^2(t) \rangle^{1/2}$, one obtains $s$ and $\dot s$ at
$t=t_{\rm dec}$.  As shown in ref.\,\cite{Buchmuller:2014rfa}, the
classical waterfall field reaches the local, inflaton-dependent
minimum soon after the decoherence time,
\begin{align}
s_{\rm min}^2(\varphi)=2\xi -(\lambda^2/g^2) \varphi^2 \ , 
\end{align}
and, together with the inflaton field, it reaches the global minimum
after a large number of $e$-folds.

\begin{figure}[t]
  \begin{center}
    \includegraphics[scale=0.6]{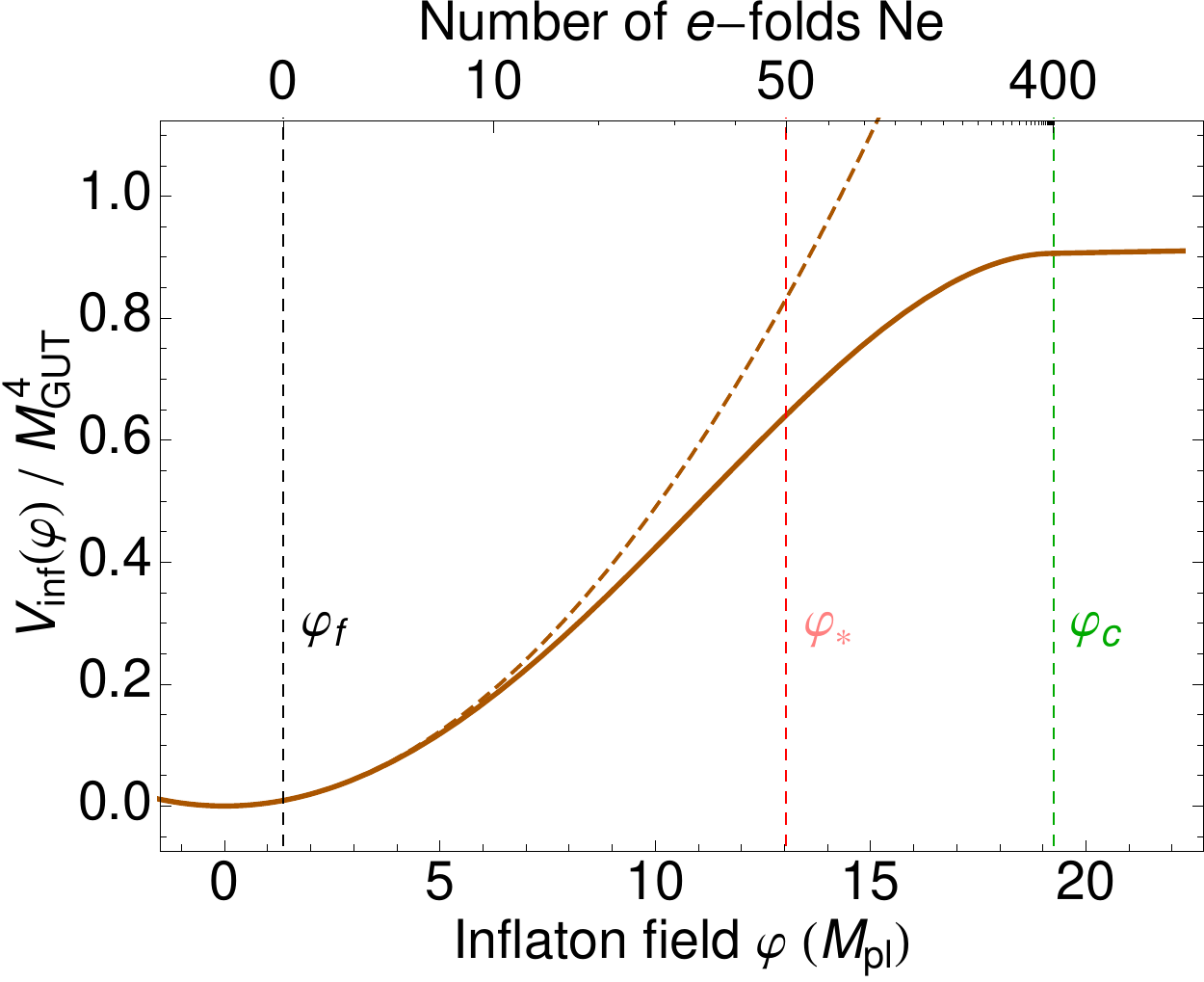}
  \end{center}
  \caption{Effective inflaton potential in subcritical hybrid
    inflation (solid line) normalized to $M_{\rm GUT}^4$, with $M_{\rm
      GUT} = 2\times 10^{16}\,\mathrm{GeV}$.  For reference, a
    quadratic potential is shown (dashed line). $\varphi_c$,
    $\varphi_*$ and $\varphi_f$ are the inflaton field values at the
    beginning of the waterfall transition, beginning and end of the
    last 50 $e$-folds of inflation. Parameters: $\bar{\lambda} =
    7\times 10^{-4}$, $M_{\rm inf}=1.95\times 10^{16}\,\mathrm{GeV}$.
    (See also fig.\,3 in ref.\,\cite{Buchmuller:2014rfa}).}
  \label{fig:potential}
\end{figure}

On the inflationary trajectory, the inflaton potential takes a simple
form,
\begin{align}\label{Vinf}
V_{\rm inf}(\varphi) &=
V(\varphi,s_{\rm min}(\varphi)) \nonumber\\
&= g^2 \xi^2\ \frac{\varphi^2}{\varphi^2_c}
\left(1-\frac{1}{2}\frac{\varphi^2}{\varphi^2_c}\right) \ , \quad
\varphi \leq \varphi_c \ .
\end{align}
For small $\varphi$, the potential is quadratic, and as $\varphi$
approaches $\varphi_c$, the potential reaches the plateau $g^2
\xi^2/2$. Fig.\,\ref{fig:potential} shows the potential for a certain
choice of parameters.  As we shall see in the following sections, in
the relevant parameter range the predictions for $n_s$ and $r$ only
depend on the potential \eqref{Vinf}. The initial conditions, in
particular the initial value of $\varphi$ and the tachyonic growth of
the waterfall field only affect the total number of $e$-folds and the
formation of cosmic strings.

\section{III. cosmological observables}

\begin{figure*}[t]
  \begin{center}
    \includegraphics[scale=0.6]{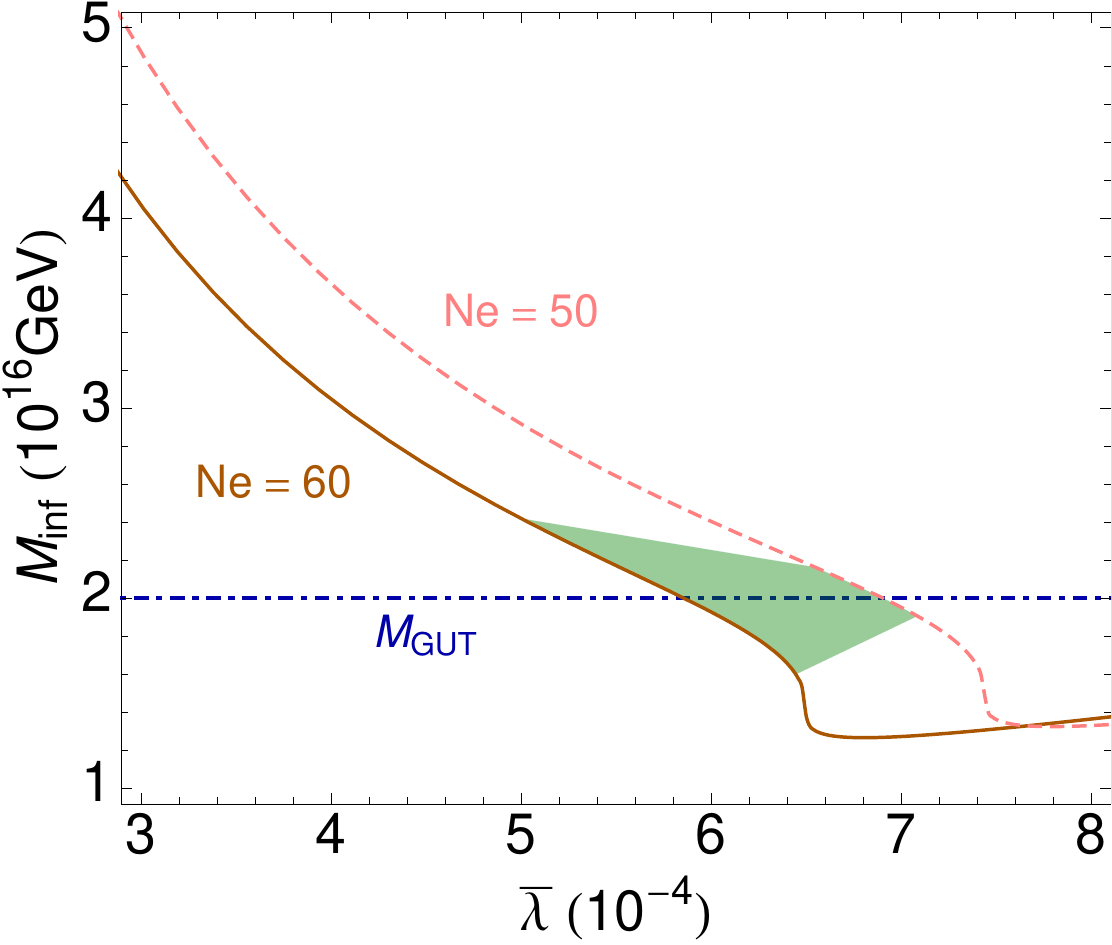}\hspace*{1cm}
    \includegraphics[scale=0.6]{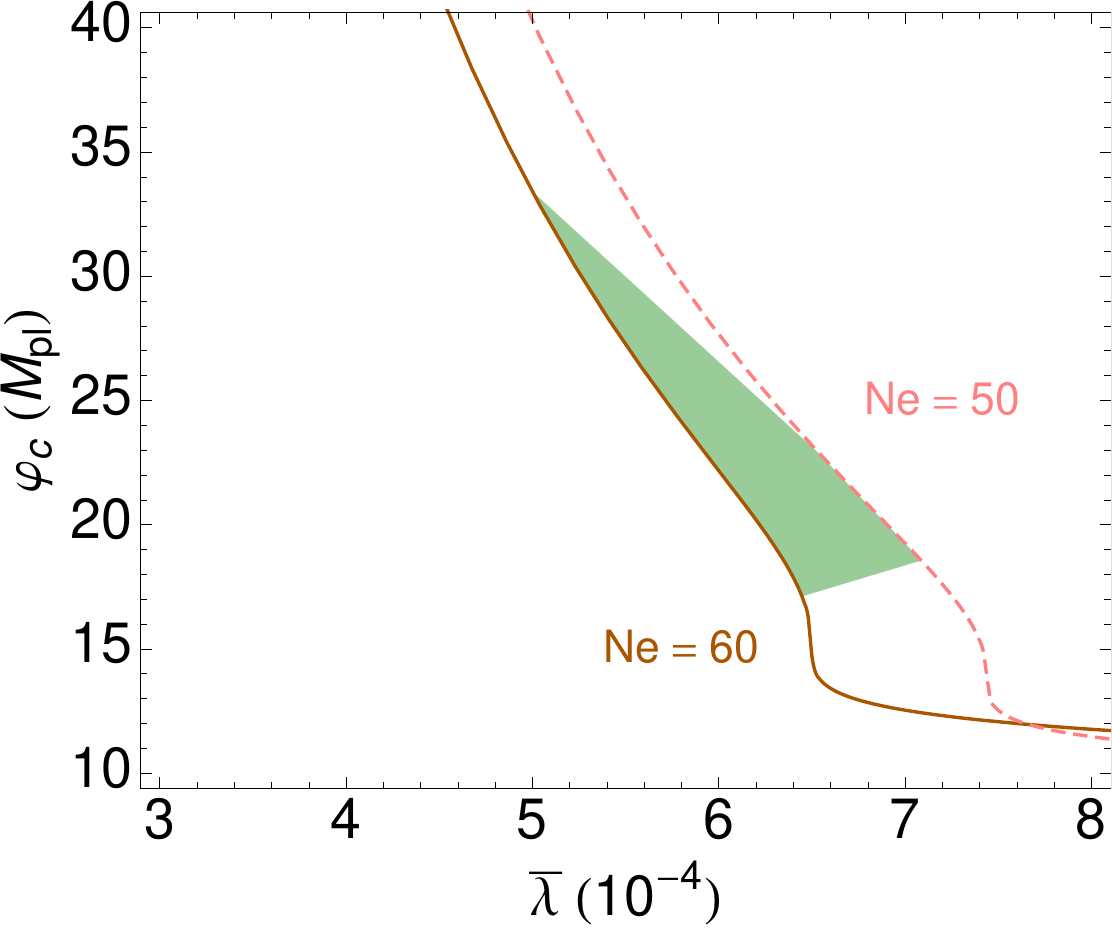}
  \end{center}
  \caption{Lines in the $\bar{\lambda}-M_{\rm inf}$ plane (left) and
    the $\bar{\lambda}-\varphi_c$ plane (right), which are determined
    by the measured amplitude of the scalar power spectrum for
    $N_e=60$ and $50$. %, respectively.
    The shaded regions are allowed by the bound on $n_s$ and $r$
    obtained from the Planck data.  $M_{\rm GUT} = 2\times
    10^{16}\,\mathrm{GeV}$.}
  \label{fig:Minf_phic}
\end{figure*}

In this section we analyse the implications of the constraints on the
cosmological observables $n_s$ and $r$ by the Planck data on the
parameters of the inflaton potential \eqref{Vinf}. Obviously, the
potential only depends on two parameters, which can be chosen as
\begin{align}
M_{\rm inf} = \left(\frac{g\xi}{\sqrt{2}}\right)^{1/2} \ , \quad 
\bar{\lambda} = \frac{\lambda}{\sqrt{g}} \ ,
\end{align} 
where we have used $\varphi_c^2 = 2\sqrt{2}M_{\rm
  inf}^2/\bar{\lambda}^2$.  Then the energy density of the plateau is
given by $V_{\rm inf}(\varphi_c)=M_{\rm inf}^4\,$.

Scalar spectral index and tensor-to-scalar ratio are conveniently
expressed in terms of the slow-roll parameters of the inflaton
potential,
\begin{align}
\epsilon(\varphi) = 
\frac{1}{2} \left(\frac{V_{\rm inf}'}{V_{\rm inf}}\right)^2 \ ,
\quad
\eta(\varphi) = \frac{V_{\rm inf}''}{V_{\rm inf}} \ ,
\end{align}
where the superscript `prime' denotes the derivative with respect to
$\varphi$, and we have set the Planck mass $M_{\rm pl}=1$. Inflation
ends at $\varphi=\varphi_f$ which is defined by ${\rm
  max}\{\epsilon(\varphi_f),\, |\eta(\varphi_f)|\}=1$.  The number of
$e$-folds between $t_*$ and $t_f$ can then be expressed as
\begin{align}
N_e=\int^{t_f}_{t_*} dt\, H =
 \int^{\varphi_*}_{\varphi_f}d\varphi
 \frac{1}{\sqrt{2\epsilon(\varphi)}} \ ,
\end{align}
where $\varphi_*=\varphi(t_*)$. Solving this equation, one obtains
$\varphi_*$ in terms of $N_e$,
\begin{align}\label{expansion}
\varphi_*^2=4N_e+2 -\sum_{n\ge 1} \frac{a_n}{\varphi_c^{2n}} \ ,
\end{align}
where $a_1=4(N_e^2+N_e+1)$, $a_2=(2/3)(2N_e^2-3)(2N_e+3)$,
$a_3=-(4/3)(N_e^4+2N_e^3+6N_e^2-3), \ldots$. The first 3 terms in the
expansion \eqref{expansion} yield $\varphi_*$ to sufficient accuracy.
Together with the standard expressions for $n_s$ and $r$,
\begin{align}
n_s=1+2\eta_*-6\epsilon_*\ , \quad 
r = 16 \epsilon_* \ ,
\end{align}
where $\epsilon_*=\epsilon(\varphi_*)$ and
$\eta_*=\eta(\varphi_*)$, this yields $n_s$ and $r$ for a given number
of $e$-folds $N_e$.

Finally, a crucial observable is the amplitude of the scalar power
spectrum $A_s$,
\begin{eqnarray}
A_s=\frac{V_{\rm inf}(\varphi_*)}{24\pi^2\epsilon_*} \ , 
\label{A_s}
\end{eqnarray}
which is determined as $A_s=2.196^{+0.051}_{-0.060}\times 10^{-9}$ at
$68$\% CL~\cite{Ade:2013zuv} from the combined data sets of the Planck
and WMAP collaborations. Imposing the central value of $A_s$ as
constraint yields a line in the $\bar{\lambda} - M_{\rm inf}$ plane
for a given number of $e$-folds. The result is shown in
fig.\,\ref{fig:Minf_phic} for $N_e = 60$ and $50$. The shaded region
is consistent with the constraints \eqref{2sigma} of the Planck data
on $n_s$ and $r$. The energy scale of the plateau is rather precisely
determined,
\begin{align}
1.6\ (1.9) \,\le 
\left(\frac{M_{\rm inf}}{10^{16}\,{\rm GeV}}\right) \le\, 
2.4\ (2.2) \ , 
\label{xiN60}
\end{align}
for $N_e=60\,(50)$. It is very remarkable how accurately the energy
scale $M_{\rm inf}$ of the plateau agrees with the energy scale
$M_{\rm GUT}$ of gauge coupling unification in the supersymmetric
standard model. For comparison, fig.\,\ref{fig:Minf_phic} also shows
the allowed region in the $\bar{\lambda} - \varphi_c$ plane. The
allowed values of $\varphi_c$, and also $\varphi_*$ are
super-Planckian, similar to chaotic inflation.\footnote{Note that for
  $\bar{\lambda} \gtrsim 10^{-4}$ the treatment of the initial
  tachyonic growth of the waterfall field is consistent, while
  $\bar{\lambda} \lesssim 10^{-3}$ is small enough to allow for 60
  $e$-folds below the critical point~\cite{Buchmuller:2014rfa}.}

\begin{figure}[t]
  \begin{center}
    \includegraphics[scale=0.6]{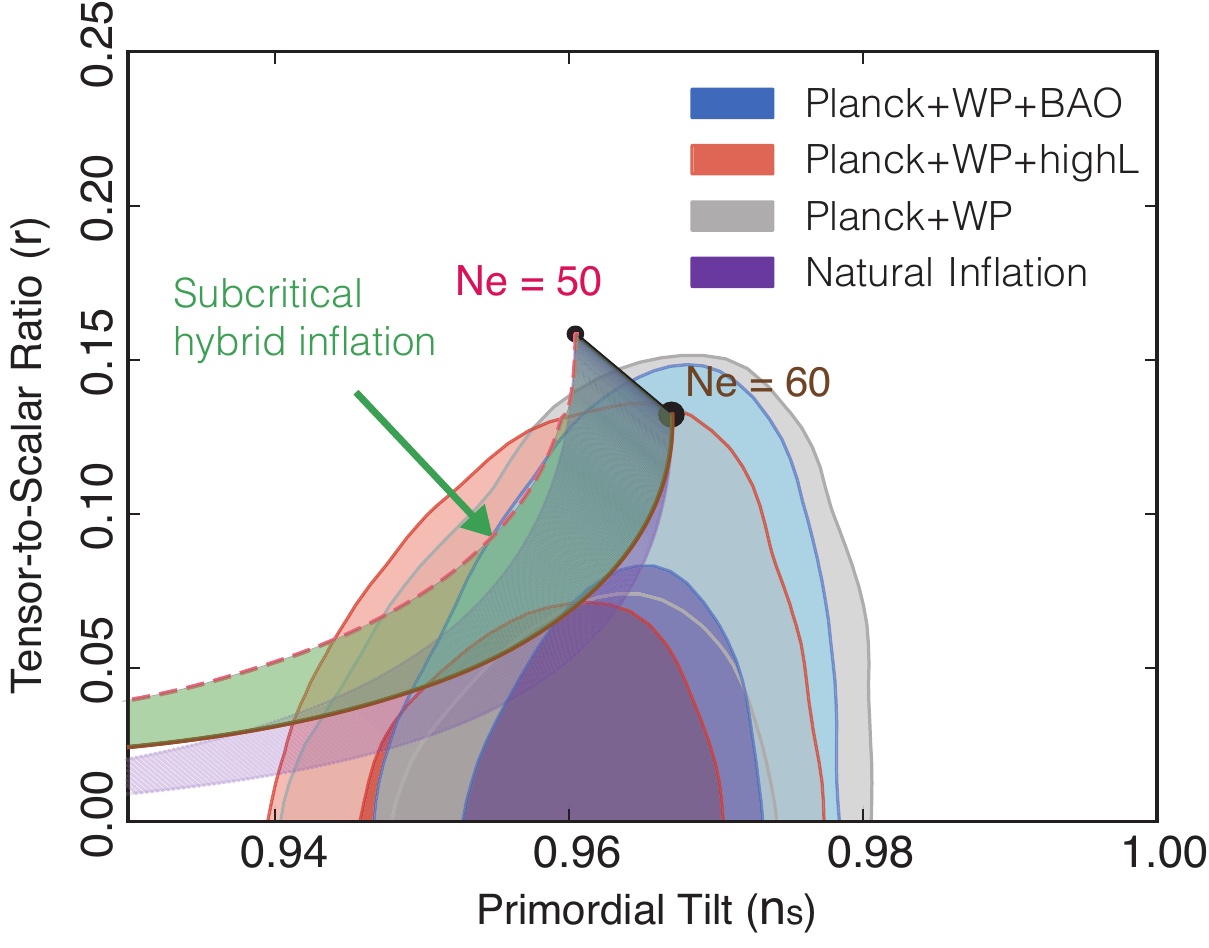}
  \end{center}
  \caption{$68$\% and $95$\% CL regions in the $n_s-r$ plane from
    Planck data in combination with other data sets compared to
    natural inflation, as given in ref.\,\cite{Ade:2013uln}, and
    subcritical hybrid inflation (green).}
  \label{fig:nsr}
\end{figure}

Varying $\bar{\lambda}$ yields a line also in the $r - n_s$ plane for
a given number of $e$-folds. In fig.\,\ref{fig:nsr} the result is
compared with various constraints from CMB data and the prediction of
natural inflation \cite{Ade:2013uln}. As one can see, subcritical
hybrid inflation and natural inflation
\cite{Freese:1990rb,Freese:2014nla} yield qualitatively similar
predictions. This is not surprising, given the similarity of the
potential \eqref{Vinf} to a cosine-potential.\footnote{A similar
  potential can be obtained in chaotic inflation with nonminimal
  coupling to gravity \cite{Linde:2011nh}. For a recent discussion of
  universality classes for models of inflation, see
  ref.\,\cite{Binetruy:2014zya}.}  The interpretation, however, is
very different. In natural inflaton the band is obtained by varying a
super-Planckian axion decay constant or, as in aligned two-axion
models \cite{Kim:2004rp}, the ratio of sub-Planckian decay
constants. On the contrary, in subcritical hybrid inflation different
points of the band correspond to different values of the ratio of a
small Yukawa coupling and a gauge coupling $\mathcal{O}(1)$.  The
lower bound from the Planck data on the spectral index $n_s$ implies
the lower bound on the tensor-to-scalar ratio $r > 0.049\, (0.085)$
for $60\, (50)$ $e$-folds before the end of inflation.

Let us finally comment on the formation of cosmic strings in
subcritical hybrid inflation.  Cosmic strings are produced during the
tachyonic growth of the waterfall field, which spontaneously breaks
the U(1) symmetry. The initial average distance of the cosmic strings
can be estimated as~\cite{Buchmuller:2014rfa}
\begin{align}\label{estimate}
d_{\rm cs}(t^{\rm loc}_{\rm sp})\sim 
k_b^{-1}(t)a(t)|_{t=t^{\rm loc}_{\rm sp}} =
\mathcal{O}\left(\frac{1}{H_c}\right) \ ,
\end{align}
where $a(t)$ is the scale factor and $t^{\rm loc}_{\rm sp}$ is the
spinodal time at which $s(t)$ reaches the local, inflaton-dependent
minimum $s_{\rm min}(\varphi)$.  Between $t^{\rm loc}_{\rm sp}$ and
$t_*$, the beginning of the last 50--60 $e$-folds, the scale factor
grows by $\Delta N_e$ $e$-folds whereas the Hubble parameter remains
almost constant, $H_* \sim H_c$, which yields for the average string
separation at $t_*$,
\begin{align}
d_{\rm cs}(t_*) \sim e^{\Delta N_e} \frac{1}{H_*} \ .
\end{align}
The smallest value of $\Delta N_e$ is obtained for the largest
coupling $\bar{\lambda}_{\rm max} = 7\times 10^{-4}$: $\Delta N_e^{\rm
  min} \simeq 380$ (see figs.\,\ref{fig:potential} and
\ref{fig:Minf_phic}). 
During the final $50$--$60$ $e$-folds the
horizon at $t_*$ is blown up to $1/H_0$, the size of the present
universe. We thus obtain the lower bound on the average cosmic string
distance
\begin{align}
d_{\rm cs}(t_0) > e^{380} \frac{1}{H_0} \ .
\end{align}
Hence, cosmic strings are unobservable in subcritical hybrid inflation
for parameters consistent with the Planck data.

\section{IV. conclusions}

We have studied subcritical hybrid inflation, which occurs in
supersymmetric D-term inflation for small couplings of the inflaton to
matter. The effective inflaton potential interpolates between a
quadratic potential at small field values and a plateau at large field
values. It is characterized by two parameters, the energy scale of the
plateau, and the critical value of the inflaton field, at which the
plateau is reached.

The model can accommodate the Planck data very well, and it is
striking how accurately the energy scale $M_{\rm inf}$ of inflation
agrees with the scale $M_{\rm GUT}$ of gauge coupling unification in
the supersymmetric standard model. This reopens the question on the
possible connection between grand unification and inflation.

The predictions for the scalar spectral index and tensor-to-scalar
ratio are qualitatively similar to those from natural
inflation. Quantitatively, however, the predicted values for the
tensor-to-scalar ratio are larger and one obtains the lower bounds
$r>0.049\,(0.085)$ for $60\,(50)$ $e$-folds before the end of
inflation, which is in reach of upcoming experiments.\\
  
\noindent
{\it Acknowledgements}\\
We are grateful to Valerie Domcke and Kai Schmitz for valuable
discussions and support.  We also thank Rose Lerner, Alexander
Westphal and Clemens Wieck for helpful comments on the manuscript.
This work has been supported in part by the German Science Foundation
(DFG) within the Collaborative Research Center 676 ``Particles,
Strings and the Early Universe''.

\end{document}